# Fundamental design paradigms for systems of three interacting magnetic nanodiscs

D. M. Forrester[1], K. E. Kürten[1,2], F. V. Kusmartsev[1]

[1]Department of Physics, Loughborough University, Loughborough, LE11 3TU, UK

[2]Faculty of Physics, University of Vienna, 5, Boltzmanngasse, A-1090, Vienna, Austria

The magnetic properties of a system of three interacting magnetic elliptical discs are examined. For the various levels of uniaxial anisotropy investigated a complicated series of phase transitions exist. These are marked by the critical lines of stability (CLS) that are demonstrated in an applied magnetic field plane diagram.

As a magnetic field gradient is applied to a magnetic multilayer (MML) of nanodiscs there emerges a series of reversible (RPC) and/or discontinuous phase changes (DPC's) in the magnetization that are associated with the response of the magnetic moments (MM's) to the field. With understanding of the magnetic field strengths at which these phase changes occur, the precise targeting of the MML's response characteristics to an applied magnetic field (AMF) can be done. For example, the torque invoked upon a group of nanomagnets by a small AMF may be used in signal transduction therapy, whereby the mechanical stimulation of the cellular membrane of a cancer cell results in a type of cell death[1]. The torque to do this need not be that achieved at the saturation field and may occur around a DPC of the MM orientation. Thus, we examine the system of three interacting MML's and show the phases associated with magnetization reversals.

For the MML stack of three nanomagnets (each has the same volume (V), saturation magnetization ($M_S$), and uniaxial anisotropy constant (D)) interspersed by insulating layers we apply a quasi-static analysis (see [2, 7-10] for comparison). The easy axes of the magnetic layers (ML's) are taken to be parallel to one another and the AMF ($\mathbf{h}_a = h_a(\cos\beta, \sin\beta)$) is applied in the $x-y$ plane of the magnetic disc at an angle $\beta$ to the easy axes.

$$\varepsilon = -\mu_0 V M_S^2 \left( J \sum_{i=1}^{2} \mathbf{m}_i \cdot \mathbf{m}_{i+1} - D \sum_{i=1}^{3} (\mathbf{m}_i \cdot \hat{\mathbf{e}}_y)^2 + \sum_{i=1}^{3} \mathbf{h}_a \cdot \mathbf{m}_i \right) \quad (1)$$

Assuming that the ML's are thin enough that the magnetization moves in the $x-y$ plane too: $\mathbf{m}_i = (\cos\varphi_i, \sin\varphi_i, 0)$. Equation 1 can be simplified for numerical calculations by writing $J^* = \mu_0 V M_S^2 J$, $D^* = \mu_0 V M_S^2 D$, $H^* = \mu_0 V M_S^2 h_a$, $E = \varepsilon/|J^*|$, $K = D^*/|J^*|$, $H = H^*/|J^*|$ and $\eta = J^*/|J^*|$. Thus, we are left with the dimensionless energy,

$$E = -\eta \sum_{i=1}^{2} \cos(\varphi_i - \varphi_{i+1}) + \frac{K}{2} \sum_{i=1}^{3} \sin^2 \varphi_i - H \sum_{i=1}^{3} \cos(\varphi_i - \beta) \quad (2)$$

The constant $\eta$ is -1 (+1) for anti-parallel (parallel) coupling of the layers. The magnetization angles in Eq. 2 are found by solving the Landau Lifshitz Gilbert equation (LLGE), $-\frac{\partial \mathbf{m}_i}{\partial \tau} = \mathbf{m}_i \times \mathbf{h}_{eff} + \alpha \mathbf{m}_i \times \mathbf{m}_i \times \mathbf{h}_{eff}$. Here $\mathbf{h}_{eff}$ is an effective field given by $\partial E/\partial \mathbf{m}_i$. The dimensionless time is $\tau = \gamma M_S t$ with gyromagnetic ratio $\gamma$. Under the assumptions that the layers are thin, that $H_z = 0$ and that we can to a good approximation take a quasi-static approach, we develop three coupled non-linear equations, derived from (2) and LLGE, which are equivalent to $\partial E / \partial \varphi_i = 0$ [12]. The magnetization angles are attained by numerical solution of these equations using Newton and gradient descent procedures. They are found to be related to energy minima when the eigenvalues of the Hessian matrix are greater than zero. The net magnetization for these orientations of $\varphi_i$ is shown in Fig. 1 when K=1.7 and the AMF is at $\beta = 30°$ to the easy axes. When each layer has the same magnetization orientation, we describe the system as having parallel (P) alignment. In Fig. 1 the layers have this P alignment beyond the critical point (CP) C5 (on branch B1). Figure 1 shows the magnetization in the field direction at various values of AMF and it is found by $M = (1/3) \partial E / \partial H$.

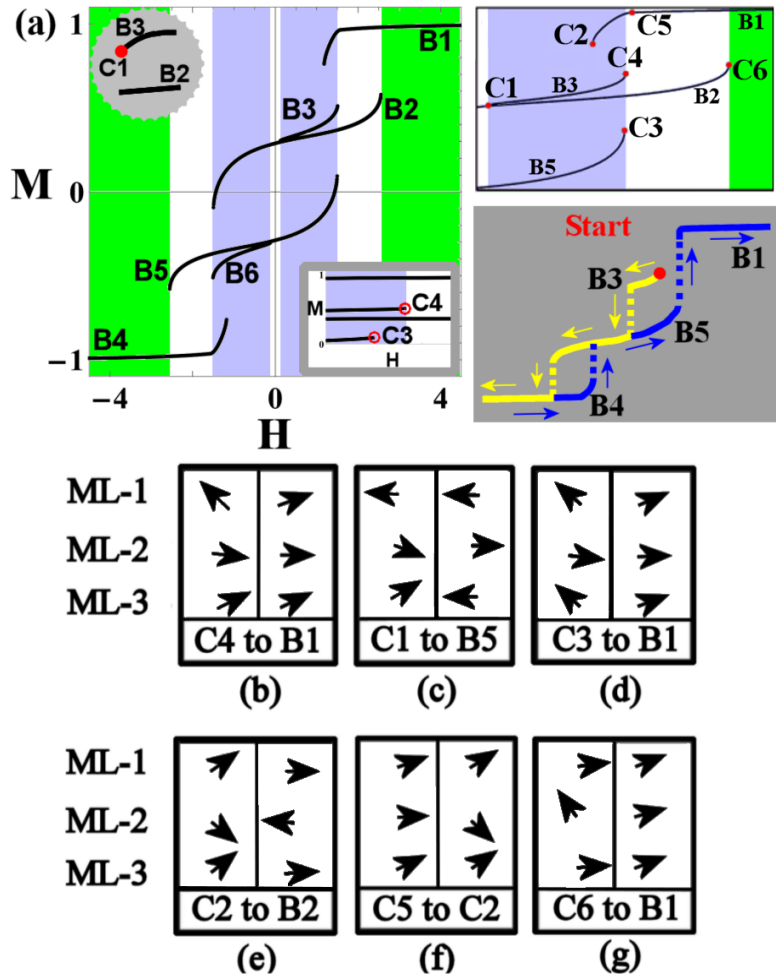

FIG. 1. The normalised net moment projected along the direction of the field as a function of the field. The magnetization angles ($\varphi_i$) in the three ML's are shown for $K = 1.7$ and $\beta = 30^\circ$. There exist six branches of magnetization, B1-B6. A gradient descent method of iterative analysis shows the hysteresis path that occurs when $\varphi_1 \neq \varphi_2 \neq \varphi_3 \neq \varphi_1$ is the starting configuration (bottom right illustration in (a)), i.e. beginning on B3. Blue (yellow) lines and arrows show the forward (reverse) evolution of the magnetization. Metastable branch B3 may be realised by a rapid cooling technique. The repetition of the fast cooling from high temperatures at different AMF strengths may drive the system to settle in the different valleys of the energy landscape (see Ref. 15). The top right illustration in (a) shows the six CP's (C1-C6) in the positive quadrant of the M-H plot at which there is a change of phase. The green areas (e.g. $H > C6$) show $\varphi_1 = \varphi_2 = \varphi_3$, exclusively. The white areas (e.g. $C4 < H < C6$) in the plots are where no perfectly non-parallel configurations (PNPC's) can appear. The light purple areas represent the range of H in which there exists the possibility of the appearance of the PNPC's that correspond to $\varphi_1 \neq \varphi_2 \neq \varphi_3 \neq \varphi_1$. CP's C1-C4 and C6 are characterised by Barkhausen jumps, whereas C5 marks the point along B1 when $\varphi_1 = \varphi_2 = \varphi_3 = 14.66^\circ$. As the AMF increases in strength $\varphi_i$ tends to the field angle. The CP's C3 and C4 are offset from one another by $H \approx 0.01$ (bottom inset of the main M-H in (a)). The top inset magnification shows that B3 does not intercept B2. (b) The transition from C4 ($\varphi_1 \neq \varphi_2 \neq \varphi_3 \neq \varphi_1$) to B1 at $H = 1.5097$ (at this point on B1, $\varphi_i$ are not quite parallel. As H increases in strength $\varphi_i$ will transition into a P configuration with a second order phase transition, marked by a CP such as C5). The top vector in each column represents the magnetization vector in the top ML,

*depicted as ML-1. Likewise, the middle and bottom layers are denoted by ML-2 and ML-3, respectively. The first column is the orientation of the magnetization vectors at C4. The second column is after the system has switched to B1. (c) The transition from C1 to B5 at $H = 0.1172$. (d) C3 to B1 at $H = 1.4978$. (e) C2 to B2 at $H = 1.1753$. (f) C5 to C2 at $H = 1.5739$ to $H = 1.1753$, respectively. (g) C6 to B1 at $H = 2.5563$.*

The energy of the system is shown in Fig. 2 for the same values of anisotropy and AMF as in Fig. 1. Branches B1 to B6 in Fig. 2 lead to a DPC at the CP's C1-C4 and C6. These mark first order phase transitions which are also illustrated in Fig. 3. CP C5 gives rise to a RPC along B1. The stability of the phases depends on the system history, i.e. from which field strength and from which orientation of $(\varphi_1, \varphi_2, \varphi_3)$ the present phase has originated from.

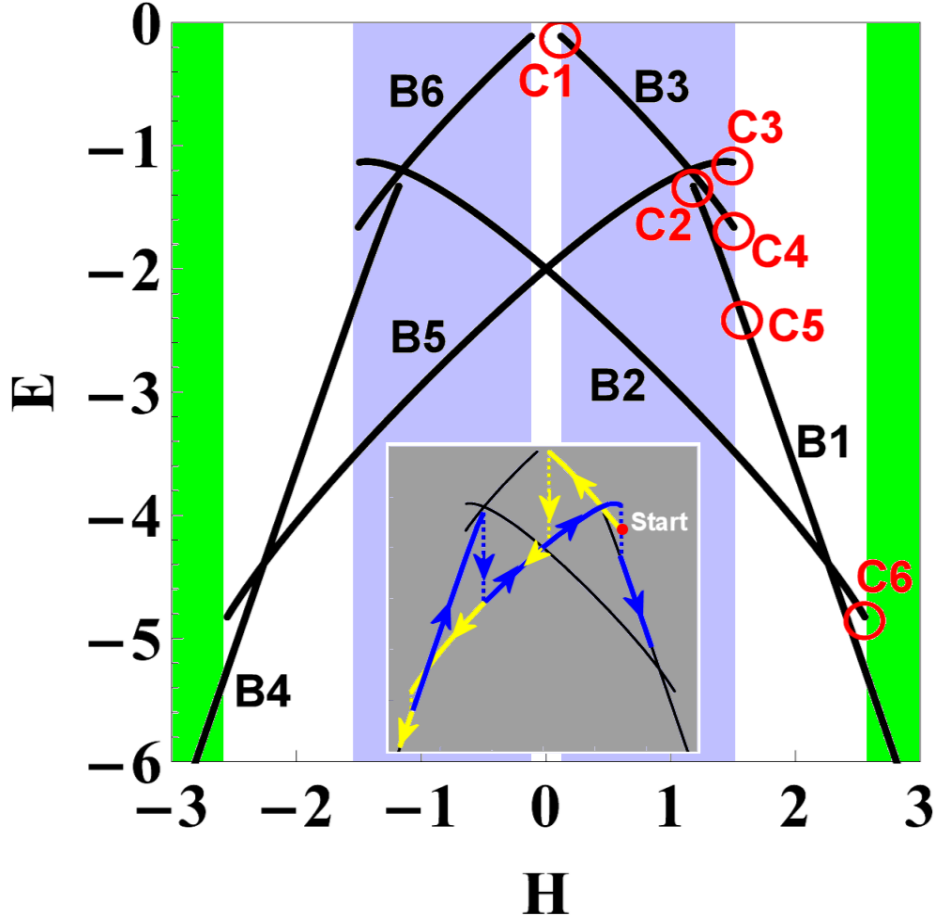

FIG. 2. The energy (E) as a function of the AMF (H) for the different stable states described in Fig.1. These stable states are associated with the branches B1-B6. The local energy minima correspond to MM configurations which are found by inserting $\varphi_i$ (the same $\{\varphi_i\}$ as in Fig. 1) into the energy equation, Eq. 2. The orientations of the MM's at the critical points here are the same as in Fig. 1 and can be seen in Fig. 1 (b)-(g) for a comparison. The CP's all lie on the end of these branches, except for C5, which lies on B1. CP's C1, C2, C3, C4 and C6 are the points where first order phase transitions occur and there is a discontinuity in the first derivative of the energy with respect to the AMF. C5 is where a second order transition occurs between a non-saturated and saturated state, i.e. near C5 the distinction between phases becomes almost non-existent. The inset shows the hysteresis path shown in Fig. 1 as it appears through the energy branches.

The critical line marking the transition to P alignment (Fig. 3 and the left hand phase diagrams in 4 (c), (d), and (e)) of the MM's in the three layers is given by,

$$(H_x/K)^{2/3} = \left((1-F^2)(1-F^2+(3\eta/K))^2\right)^{1/3} \qquad (3)$$

Where $F(H_y) = (s-(H_y/2K))^{1/3} - (s+(H_y/2K))^{1/3}$ and $s = \sqrt{(-\eta/K)^3 + (H_y/2K)^2}$. For the uncoupled system ($\eta=0$) this reduces to the Stoner-Wohlfarth astroid[10] valid for a single ML, $(H_x/K)^{2/3} + (H_y/K)^{2/3} = 1$.

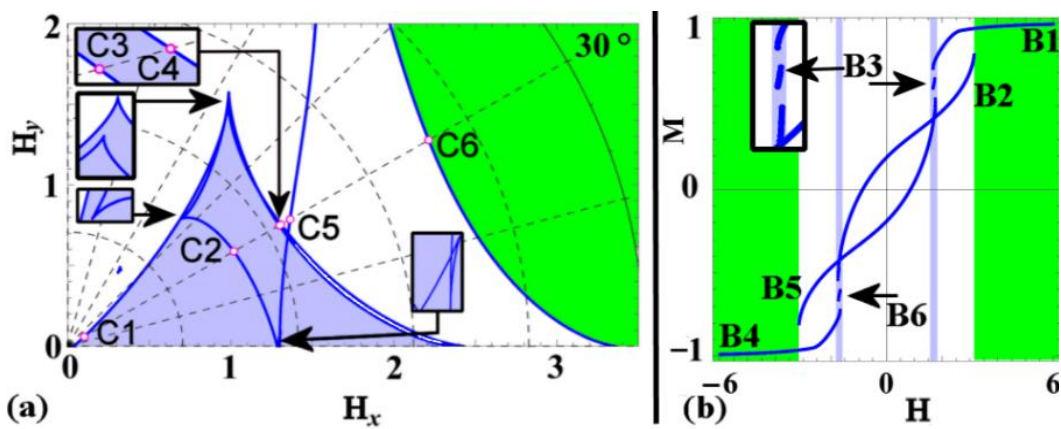

*FIG. 3. (a) The CLS (blue lines online) are associated with local energy minima where some minima vanish or bifurcate into other ones. Here the critical lines are illustrated for a system of three interacting magnetic discs with anisotropy constant $K=1.7$. For the situation whereby the AMF is applied at $30°$ to the easy axes of the layers, the critical points are outlined by the pink circles denoted by C1-C6. The line going through these points at $30°$ correlates with the illustrative cases in Fig. 1 and 2 that show the response of the system to an AMF at this angle. Here we show the CLS in the $H_x - H_y$ plane for $\beta=0$ to $90°$. Here,*

$(H_x, H_y) = H(\cos\beta, \sin\beta)$. *The three other quadrants of the diagram are symmetric to this first quadrant depiction. Beyond the line that C6 lies on (green area) the magnetization vectors in the discs align in parallel with each other. The white areas in the phase diagram are when there is the possibility that $\varphi_i$ are all equal or $\varphi_1 = \varphi_3 \neq \varphi_2$ (i.e. configurations giving rise to branches B2 or B5 in Fig. 1 and 2 which have an anti-parallel configuration of $\varphi_i$ at $H = 0$ ).The light purple areas are where there exist the PNPC's that have magnetization angles that are all different from each other. When the AMF is at an angle larger than $\beta \approx 58°$, in this first quadrant of the phase diagram, the possibility of the phase with $\varphi_1 \neq \varphi_2 \neq \varphi_3 \neq \varphi_1$ disappears (this occurs at the peak of the topmost cusp of the purple region at $(H_x, H_y) \approx (0.99, 1.57)$ ). The branches that were discussed in Fig. 1 and Fig. 2 have merged together. This can be seen to be occurring in (b), which shows the M-H plot for $\beta = 55°$. The naming convention for the branches of Fig. 1 and 2 are preserved so that it can be seen that B3 and B6 have now become very small (the light purple region). As the magnetic field angle approaches $\beta = 58°$, B1, B3 and B5 begin to merge (as do B2, B4 and B6). Thus, the purple domain in (a) and (b) vanishes, leaving only the white domain.*

Two special cases are now demonstrated for $\beta = 0$ and $90°$. In these cases the MM's in each layer can have an anti-parallel (AP) configuration (e.g. $(\varphi_1, \varphi_2, \varphi_3) = (\pi, 0, \pi)$). In the generic phase diagram (for comparisons see[11-12]) with $\beta = 90°$, Fig. 4 (a), the P to spin flop (SF) phase transition occurs at the CLS, $\pm(K+3)$. There are also SF (pink region online) to AP (red region online) transitions at $\pm(1/2)+(1/2)\sqrt{1-12K+4K^2}$ and $\pm(1/2)-(1/2)\sqrt{1-12K+4K^2}$ for $0 \leq K \leq 0.086$. This can be seen in Fig. 4 (b), in the top right plot. The pink region around H=0 is the SF phase and the red regions are AP, demonstrating that there is no remanence when $\beta = 90^o$. If the system is magnetized to $M_S$ by the applied field at $\beta \neq 90^o$, then when $H \to 0$ the magnetization reduces to a non-zero remanent magnetization. These equations for the CLS can also be applied to the CP's that lie on the x and y axes of the $H_x - H_y$ phase diagrams (see Fig. 3 and 4). Figure 4 (b) shows a more complicated generic phase diagram that exists for $\beta = 0$. A SF$\leftrightarrow$AP transition can be described by $(1/2)\pm(1/2)\sqrt{1+12K+4K^2}$ for values $0 \leq K \leq 0.4$, whereas for $0.4 \leq K \leq 1.2$ the critical line is described by $(3-K)(27K/10(K+3))^{1/3}$. Gradient descent analysis shows that the switching to the P phase ($AP \leftrightarrow P$) is history dependant and as such two possible CLS for this transition occur in Fig. 4 (b). The first of these CLS can be seen in Fig. 4 (b) to intercept the (H,K) origin. It is given by $-(1/2)\pm(1/2)\sqrt{1+12K+4K^2}$ for $K \geq 1.2$ and $H \geq 1.8$ or $H \leq -1.8$. The second of these CLS is offset from the first by $\approx \pm 1$ - i.e. $(1/2)\pm(1/2)\sqrt{1+12K+4K^2}$ for $K \geq 0.75$ and $H \geq 2.25$. In the M-H plots in Fig. 4 the dark blue lines show the forward evolution of M as H is increased from a negative value. For example, in Fig. 4 (d) and (e) the M-H plots for $\beta = 0$ have a hysteresis that has AP to P transitions in the $H > 0$ half of the diagram. System history dictates the MM configuration and an alternative transition from the other AP configuration to the P branch is also possible. This means that in the K-H plots the AP phase can extend from the illustrated area to the next CLS. Consequently, in Fig. 4 (d) and (e) the AP phase along the $H_x$ and $H_y$ axes can continue until the outer astroid-alike critical lines. The plots in Fig. 4 are for the cases when at least two of the magnetization vectors are parallel. There also exist phases for three layer interactions, seen in Fig. 1-3, where none of the magnetization vectors have the same orientation.

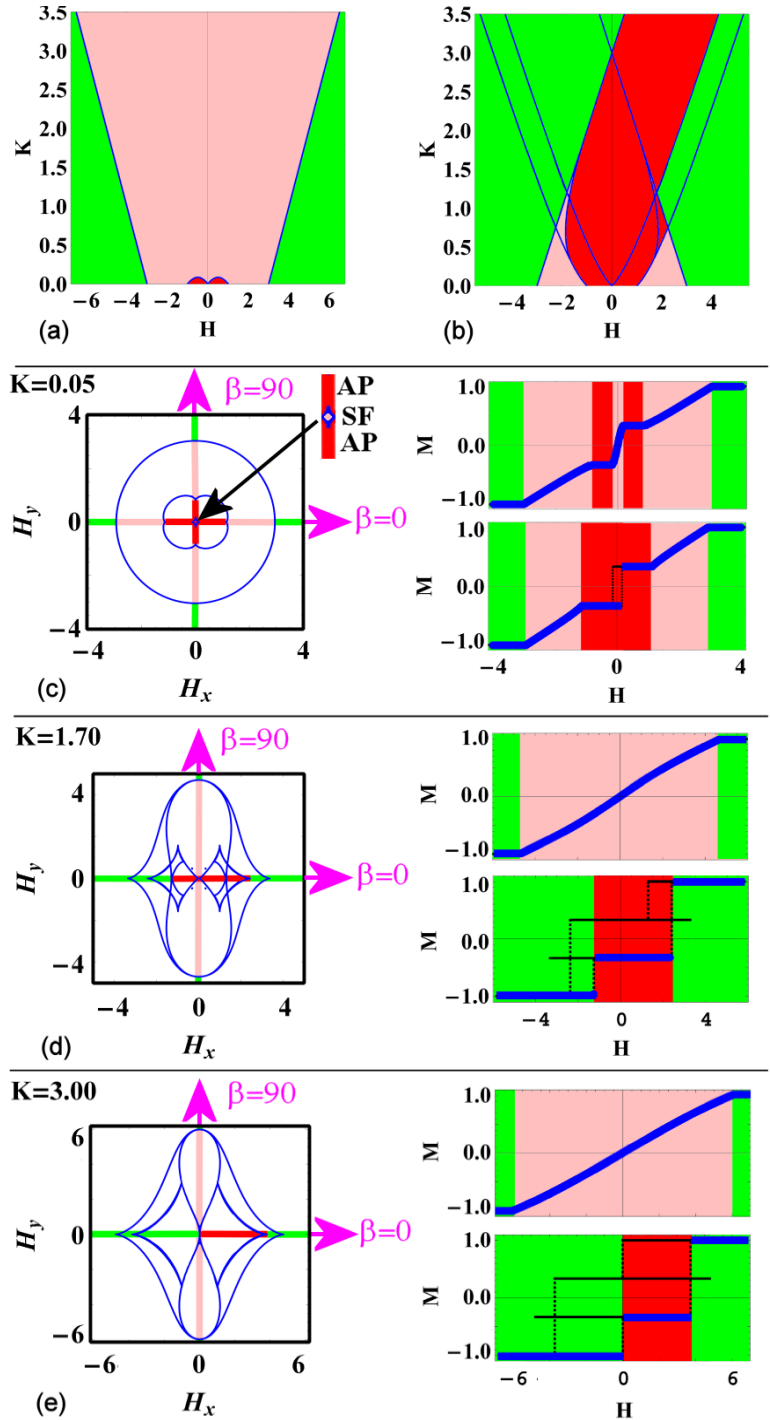

*FIG. 4. Diagrams representative of the magnetic response of three interacting ML's. (a) The CLS are shown in a generic phase diagram for K as a function of H applied*

*at $\beta = 90°$. The central area of the diagram (pink online), between the two main critical lines, is representative of a SF phase. The phases in the regions prior to and immediately after the SF phase (green online) are when the MM's in the layers are P. The small humps in the phase diagram either side of $H = 0$ (red online) are phases that have an AP origin. (b) The generic phase diagram for $\beta = 0$ (colors are as indicated in the description of (a)). The CLS themselves are symmetric around $H = 0$. The asymmetry in the coloring is because we start from a negative saturating field, P state, and start to increase H to the positive saturating field value (see Ref. 12). To see the diagrams of going from a positive saturation field to the negative one, simply perform a mirror reflection around $H = 0$. In (c) to (e) the phase diagram on the left shows the CLS that separate the P, AP and SF phases for all applied field angles at distinct values of K. On the right, the top figure is the M versus H evolution for $\beta = 90°$ and the bottom figure is for $\beta = 0$. M is the component of magnetization in the applied field direction. The thick line (blue online) shows the magnetization as H is increased from a negative strength that corresponds to a P phase.*

Understanding the magnetic response of systems of MML'S to a controlling magnetic field is crucial for designing functionalized nanomagnets. We have found that under application of a magnetic field three interacting magnetic discs undergo a series of magnetic transformations. These are observed to occur where a magnetization undergoes a Barkhausen jump or smooth second order phase transition. The three coupled ML system exhibits a phase consisting of perfectly non-parallel magnetic moments that does not exist for two layer systems. A complete analytical analysis of critical lines will follow elsewhere.

DMF thanks the EPSRC for funding under grant – EP/F005482/1